\documentclass{soups}

\usepackage{times}
\usepackage{amssymb}
\usepackage{graphicx}
\usepackage{bmpsize}
\usepackage{epsfig}
\usepackage{latexsym}
\usepackage{amsfonts}
\usepackage{eucal}
\usepackage{amsmath}
\usepackage{algorithmic}
\usepackage{algorithm}
\usepackage{stfloats}
\usepackage{fixltx2e}
\usepackage{setspace}
\usepackage[nodots]{numcompress}
\usepackage{array}
\usepackage{caption}
\usepackage{subcaption}

\begin{document}

\title{An HMM-based Multi-sensor Approach for Continuous Mobile Authentication\titlenote{This paper was published as \cite{roy2015hmm}.}}

\author{
\alignauthor
Aditi Roy, Tzipora Halevi and Nasir Memon\\
\affaddr{New York University, Polytechnic School of Engineering, Brooklyn, NY, USA}\\
\email {ar3824@nyu.edu, thalevi@nyu.edu, memon@nyu.edu}}

\maketitle

\begin{abstract}
With the increased popularity of smart phones, there is a greater
need to have a robust authentication mechanism that handles
various security threats and privacy leakages effectively.
This paper studies continuous authentication for touch interface
based mobile devices. A Hidden Markov Model (HMM) based behavioral
template training approach is presented, which does not require
training data from other subjects other than the owner of the
mobile device and can get updated with new data over time. The
gesture patterns of the user are modeled from multiple sensors -
touch, accelerometer and gyroscope data using a continuous
left-right HMM. The approach models the tap and stroke patterns of
a user since these are the basic and most frequently used
interactions on a mobile device. To evaluate the effectiveness of
the proposed method a new data set has been created from 42 users
who interacted with off-the-shelf applications on their smart
phones.
Results show that the performance of the proposed approach is
promising and potentially better than other state-of-the-art
approaches.

\end{abstract}


\section{Introduction}
\label{sec:intro}

Traditional access control mechanisms used in most smart phone
systems today, like password or lock pattern, are vulnerable to
various attacks, including password guessing and eavesdropping
(\cite{Jo12,data12}).

In search of an enhanced secure authentication mechanism, various
works explored graphical passwords \cite{DP00, DMR04, AD04, WK04,
Birget06, DMB07}. Due to the growing popularity of
touch-interfaces, research started exploring authentication
methods that exploit touch patterns \cite{SM13, van2014finger}. Sae-Bae et. al
showed that users can be authenticated  from their multi-touch
interaction through extraction of touch characteristics
\cite{SMI12, SMI12b}. However, both graphical and multi-touch
gestures have only been proposed at the login stage and therefore
do not provide protection against unauthorized access after that
stage.

Continuous authentication has been suggested as a method of
providing persistent protection to the users \cite{Aksari09,
Niinuma10}. Here, instead of authenticating a user at the time of
login, the system continuously monitors aspects of the user
behavior biometrics, like keystroke dynamics \cite{Serwadda13},
speaking pattern \cite{Woo06}, device use patterns \cite{Li13} and
touch pattern \cite{Senguard11, FAST12, Frank13, SilentSense13,
Chan14} in order to maintain authentication after login.

Touch behavior biometrics provide a promising direction for
continuously monitoring the user input on mobile devices. The
existing approaches \cite{Senguard11, FAST12, Frank13,
SilentSense13} typically employed a binary classifier trained with
the touch data from both the legitimate user as well as multiple
other users. Reported results demonstrated that such classifiers
detected the authentic user with high probability. However, in
real-life, training a system with data from other users is not
feasible.

This paper presents a Continuous HMM based Authentication System
(CHAS) on touch devices. While a basic HMM-based algorithm has
been developed before in \cite{roy2014hmm}, that method worked
only on horizontal and vertical slide data \cite{Frank13} captured
from mobile devices in constrained situation. The method has been
extended here to perform recognition on general touch-pad input
gestures, such as tap, up-down or forward-backward slide. In
addition, it provides a method for combining data from multiple
sensors, like accelerometer and gyroscope, to achieve a higher
probability of detection, without interfering with the usage model
of the device. The method offers the possibility of being trained
using only the authentic user's data and can be updated with new
data over a period of time as needed.

The proposed CHAS framework performance is demonstrated on
a newly acquired data set that allows general touch-pad input by
the user in a real-world scenario. In-depth analysis of the
proposed method using this data set shows improved authentication
accuracy can be achieved when multi-sensor data is combined with
touch information.

The key contributions of this paper are thus two-fold. 
First, the paper introduces CHAS, an improved method for HMM-based
continuous authentication, that works based on multi-sensor touch
information.
Second, the work includes a new data set that features
multi-sensor data from 42 users. Unlike the previous data set that
was used to test continuous mobile authentication,
(\cite{Frank13}) which had only limited gestures (i.e., horizontal
or vertical slides) created in controlled manner, the new data set
captures data in unconstrained situations. This study includes
extensive evaluation of the newly proposed framework on this data
set.

The rest of the paper is organized as follows. Section 2 presents
the survey to examine the need for continuous authentication on
mobile phones. The CHAS framework is introduced in Section 3. In
Section 4 the extensive experimental results are discussed and
Section 5 includes the conclusions for the work.

\section{Exploratory Study}

Previous  work was developed based on the assumption that
continuous authentication would be beneficial for the users and
that users are worried about the data on their phone.
However, no user study that the authors are aware of has been done
to support this idea. In this section we present a first attempt
to examine the need for continuous authentication by means of two
user surveys. The surveys show that participants are concerned
about the data stored on their phone and that existing locking
mechanisms are not robust enough as most of the participants
observed someone else's pin in the past. Therefore, the surveys
demonstrate the need for the development of alternatives to
pin-based or pattern based locking mechanisms.

This work includes two surveys.
The first was filled by the 47 participants at a northeastern
University \cite{Survey1}. The second was conducted online and
included 267 participants \cite{van2017draw}. Both studies were
designed to look at the mobile usage and locking patterns of the
participants.

The studies showed that most participants lock their phones (87\%
of the participants in the first study and 82\% of the
participants in the second study). However, participants in the
first study were asked if they have an application that protects
their apps/pictures/media files. Only 15\% of the participants
indicated that is the case. Most participants in this study were
worried someone would access their data in their absence (55\%).
Similarly, most participants in the second study said they had
data they would like to hide (71\%).

In addition, in the second study, participants were asked if they
ever observed the pin of another user. 73\% of the participated
did observe in the past the pin of a friend or family member, and
even a higher percentage
(79\%) knew the current pin of someone else, a finding that raises
a concern about the safety of current locking mechanisms.

The overall results can be viewed in Table \ref{SurveyTable}.
Since most participants stated they have data they deem as private
on the phone, these studies reveal the need to further continue
looking for alternative methods for mobile authentication.


\begin{table}[ht]
\begin{center}
\centering
\begin{tabular}{|c|c|c|}
\hline
   & Study 1 & Study 2 \\
%
%

\hline \hline
{No Participants}  & 47 & 267\\
\hline
{Lock the phone} & 87\%  & 82\% \\
\hline
{Use Pin or Pattern Lock} & 87\%    &81\%   \\
\hline
{Worried about Data Privacy} & 55\% &   71\% \\
\hline
{Observed someone else's pin} &  ---  & 73\% \\
\hline

\end{tabular}

\vspace{-0mm} \caption{Survey results - Mobile User Security. Most
users are using a lock mechanism and are worried about their data
privacy. Most of the participants in the second study also
observed someone else's pin in the past.} \vspace{5mm}

\label{SurveyTable}
\end{center}
\vspace{-5mm}
\end{table}

\section{Proposed HMM based Continuous Authentication System}

In this work, we assume that a malicious attacker has gained
access to a person's mobile phone. The device is either
unprotected (e.g., no PIN) or the attacker somehow knows the
authentication secret, for instance by shoulder surfing the owner.
The attacker can then perform undesirable actions with the device
violating the owner's privacy in his absence or without his
knowledge. The objective of the current work is to handle such
situations and make this kind of manipulation impossible by
analyzing touch behavior biometric of the user. To this end we
introduce a new Continuous HMM based Authentication System (CHAS)
described in the next section.

\subsection{Data Sources}

Usage of multiple sensors, including audio sensor, light sensor,
accelerometer, gyroscope, and magnetometer have been used to
provide secure access control \cite{HLM12, HMS12} in mobile
phones. Current work relies on touch, accelerometer, and gyroscope
data to create a behavior biometric model of the user for
authentication purposes.

A touch-screen based smart phone is typically operated through
touch commands such as tap, slide, pinch, and free stroke
handwriting. However, pinch, a two-finger gesture to do zooming,
is used less than 5\% of the time used by the overall gestures
\cite{SilentSense13}. Similarly, though handwriting is an
alternative input method to enter characters, it is not used very
often. Tap and slide are the basic and most frequently used
gestures. They are utilized to perform almost all of the actions
on the smart phone, such as typing using soft keyboard, unlocking
screen with PIN and navigating documents. Therefore, in the
current work, we develop a user behavior model based on these two
popular gestures.

In practice, touch gesture consists of a series of raw data input
forming a touch data sequence. For each gesture performed on the
touch-screen of the mobile device, we capture six raw features -
time stamp, vibration, rotation, pressure, size of touch, and
position. A pre-processing step is performed to normalize the
sampling rate of the data sequences, using linear interpolation
to derive a re-sampled signal with a uniform rate.

\subsection{Architecture of CHAS}

\begin{figure}[!t]
\centering
\includegraphics[width=0.8\columnwidth]{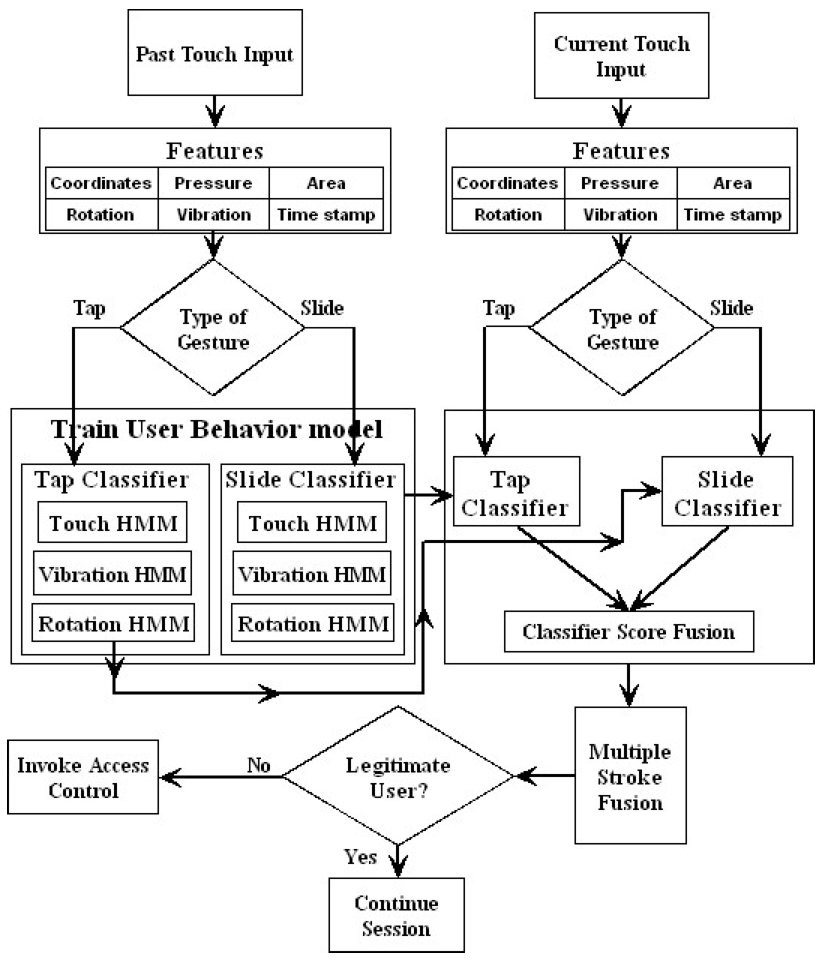}
\caption{Overview of CHAS framework}
\label{chas}
\end{figure}

To design an effective touch data based user authentication
approach, two key issues need to be addressed: 1) how to model the
user behavior from the touch data, 2) how to recognize users
according to this model. Figure \ref{chas} outlines the framework
of the proposed CHAS system. Centric to this approach is HMM based
pattern recognition algorithm that identifies the legitimate user
from touch data. As HMM was found to be better than the exiting
approaches in modeling user touch behavior \cite{roy2014hmm},
while employing only the owner's touch information, it is chosen
as the basic classifier. However, unlike the algorithm proposed in
\cite{roy2014hmm}, in which a different HMM system was needed for
each type of movement, one architecture is created for different
types of gestures in the newly proposed CHAS framework.

The proposed CHAS framework works in two steps: training and
authentication. During training, a behavioral model is created
based on the touch behavior (pressure, area, duration, position,
vibration and rotation) of the user. At the time of
authentication, the test observations are compared with the stored
behavioral model to establish the identity of the user.

After getting all the training touch data,
the behavioral model is created based on the type of gesture.
Touch-screen can capture each
user touch event and generate corresponding touch data sequence.
This touch data sequence
is considered directly as the basic input for modeling the user
behavior.
However, different sequences corresponding to different types of
touch gestures contain different spatio-temporal characteristics.
To address this issue, the proposed approach considers each type
of touch gestures separately and models them with their
corresponding sequences of raw data. Since two primary gestures
are considered here, i.e. tap and slide, two classifiers are
developed. Unlike \cite{roy2014hmm}, where separate HMM models
have been used to represent slides of different directions
(horizontal and vertical), here one single model has been employed
to represent all types of slide gestures. To achieve this, the
slide gestures are normalized to make them rotation-invariant.
Each classifier consists of three HMMs depending on the data
source. First one is based on only touch information, i.e.,
coordinate, pressure, area and time stamp. The other two are
corresponding to vibration and rotation data obtained from the
sensors.

During deployment, given a user model and some recently observed
behavior, the likelihood that the device is in the hands of the
legitimate user is computed. For both the slide and tap
classifier, the likelihood score obtained from the three HMMs are
averaged to get a combined similarity score. This value is used as
an authentication score. It may happen that the data from all the
sensors is not available or usable at a certain time. Any
continuous user identification approach using multiple sensors
should take care of such a situation. The proposed system is
flexible enough to handle subset of sensors data. It can still
work efficiently and give robust performance.

To increase the robustness of the authentication method, multiple
consecutive gestures are used for the final decision. The mean of
all the combined similarity scores is employed for this purpose.
The score is used to make an authentication decision: typically, a
threshold is used to decide whether to accept or reject the user.
The threshold can vary depending on whether the application is
security sensitive. If the input is detected to be of a legitimate
user, he can continue using the device
without any interruption. Otherwise, access control, like entering
the text or pattern based password, is invoked.

\subsection{HMM based Behavior Model}

Here the basic HMM, which is used to model the touch behavior of a
subject
as well as other sensor data input is described. HMM is considered
for modeling the touch gesture behavior of a subject since it
captures the local dynamic characteristics as well as the shape
and length of a gesture. The left-right topology is chosen with no
state skip allowed. The observable output is the trajectory
information of the gestures. The states represent the transitive
properties of the consecutive coordinates of the gesture and the
state sequence that maximizes the probability of observing the
training gestures becomes the corresponding model of a subject.

For each gesture trajectory, an M-mixture left-to-right HMM is
fitted using the Baum-Welch algorithm \cite{Rabiner93}. The
optimum number of states and mixtures of an HMM varies with the
complexity and length of gestures in the training sequences.
Five-fold cross validation principle is used to estimate the
optimal number of states and the associated HMM parameters.

Once the behavioral models for all subject classes have been
created through HMMs, authentication of the subjects is performed
by computing the log-likelihood of the input gestures using
Viterbi algorithm \cite{Rabiner93}. Along with the traditional
log-likelihood value, an additional measure named as \emph{gesture
kinematics} is computed that represents the percentage of time
spent in each state. Next, the similarity scores are derived from
the log-likelihood value and the gesture kinematics. These two
scores are combined to get the combined similarity score for
authentication. Details of the HMM based behavior model can be
found in \cite{roy2014hmm}.

\begin{figure*}[t]
\centering
\includegraphics[height=6.5 cm,width=16 cm]{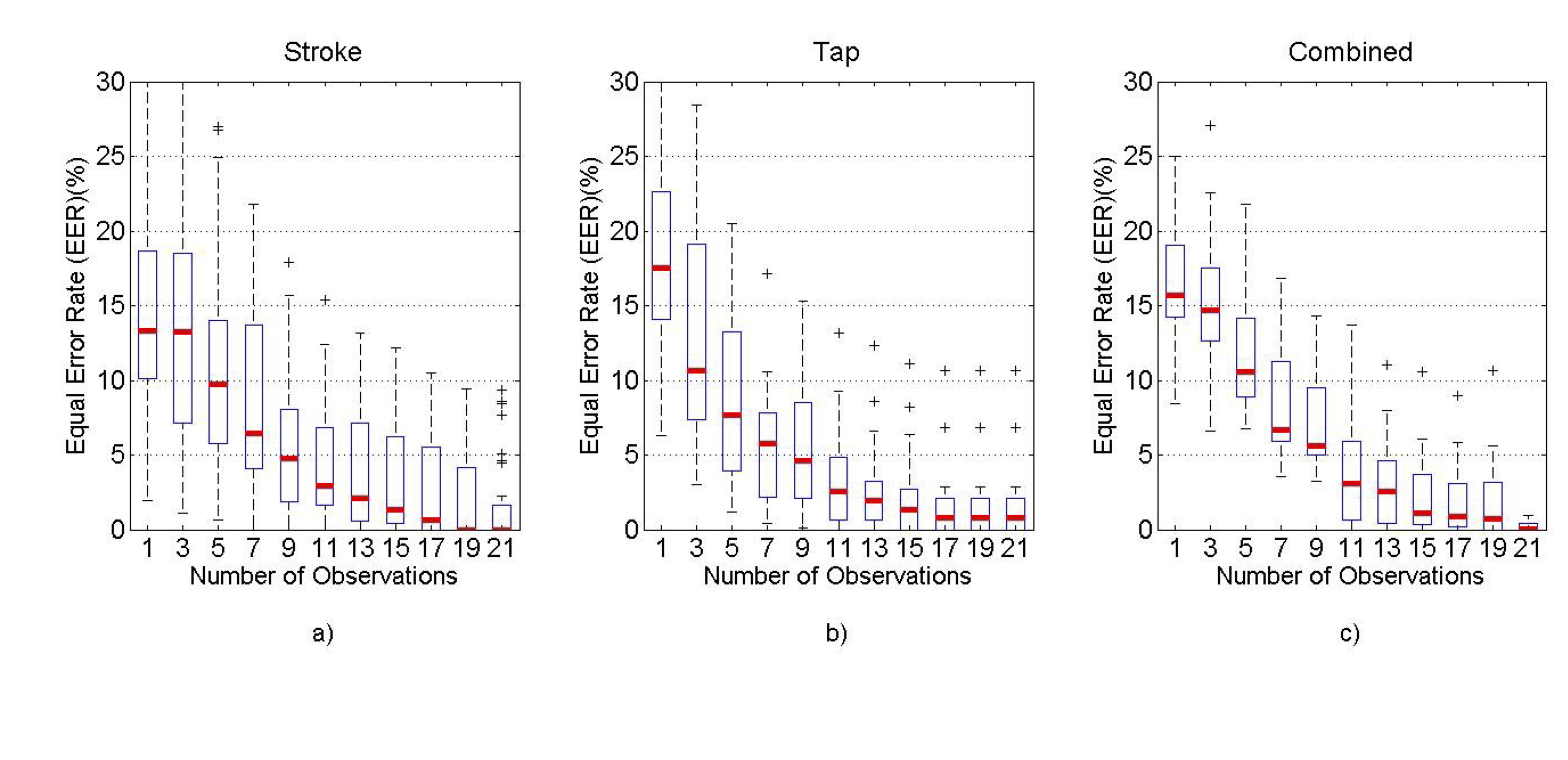}
\caption{Equal error rate variation as a function
of the number of gestures using touch-pad input} \label{wo2eer}
\end{figure*}

\section{Experimental Results and Discussion}

This section describes the performance
of the proposed CHAS framework. The authentication system is
evaluated through calculation of \emph{False Acceptance Rate
(FAR)} and \emph{False Rejection Rate (FRR)}. Since these two
error rates are inversely related (lower FAR increases the system
security while lower FRR increases its usability) \emph{Equal
Error Rate (EER)} is also measured, where the FAR is equal to the
FRR value.

Detail evaluation has been performed to test how the CHAS
framework works with different types of gestures in two possible
situations: when only touch information is used and when both
touch as well as sensor data are available. Experiments have been
performed to test the performance of the proposed system with
increasing number of consequential gestures. 

\subsection{Data Set Description}
\label{sec:Data_Set_Description}

To evaluate the CHAS framework in challenging real-life scenarios,
a new data set has been created that mimicked regular usage of
mobile devices. The users were asked to perform activities using
off-the-shelf online tools, such as reading a Wikipedia page and
filling Qualtrix questionnaires. This is in contrast to the data
set used in \cite{Frank13, roy2014hmm}, in which a
custom-application was created that enabled capturing specific
horizontal and vertical movements by the user.
The participants were asked to do a specific set of tasks using
their smart phone. They were told the goal of the study is rating
the different tasks. Those tasks included, controlling a smart
phone UI using slide touch gesture (vertical and horizontal
slide), mobile web browsing using pinch and spread touch gestures,
entering sentences using virtual touch keyboard, selecting radio
buttons by tap gesture, etc. Each task was repeated multiple times
by the same user.

Data was collected from 42 users among which 75\% of the
participants were men, 78\% of the participants belonged to the
18-24 years age group, 15\% to the 25-30 years group and 7\% to
the 30-35 years group. During data collection, users were asked to
open their mailbox and read an e-mail which had detailed
instructions about the sequence of tasks to be performed.
Afterwards, users had to open the links which pointed to the
following five tasks.
\begin{itemize}
  \item A Wikipedia page of Niagara Falls
  \item Questionnaire based on Niagara Falls Wikipedia page
  \item A Wikipedia page of NYU Poly.
  \item Questionnaire based on NYU Poly Wikipedia page.
  \item A final survey questionnaire
\end{itemize}

Every user was asked to read through each web page and then click
on a link to go to the next task. During the process, data of all
types of touch events were recorded. The data includes the
coordinates of touch, the pressure applied, the size of touch, and
the values of the accelerometer and gyroscope sensors along with
the time stamps.

The data was collected using Nexus 4 multi-touch phone with
Quad-core 1.5Ghz processor, and screen resolution $768 \times
1280$. The data capture application was developed on the Android
platform 4.4.

\subsection{Authentication with Touch Data}

\begin{figure*}[t]
\centering
\includegraphics[height=6.5 cm,width=16 cm]{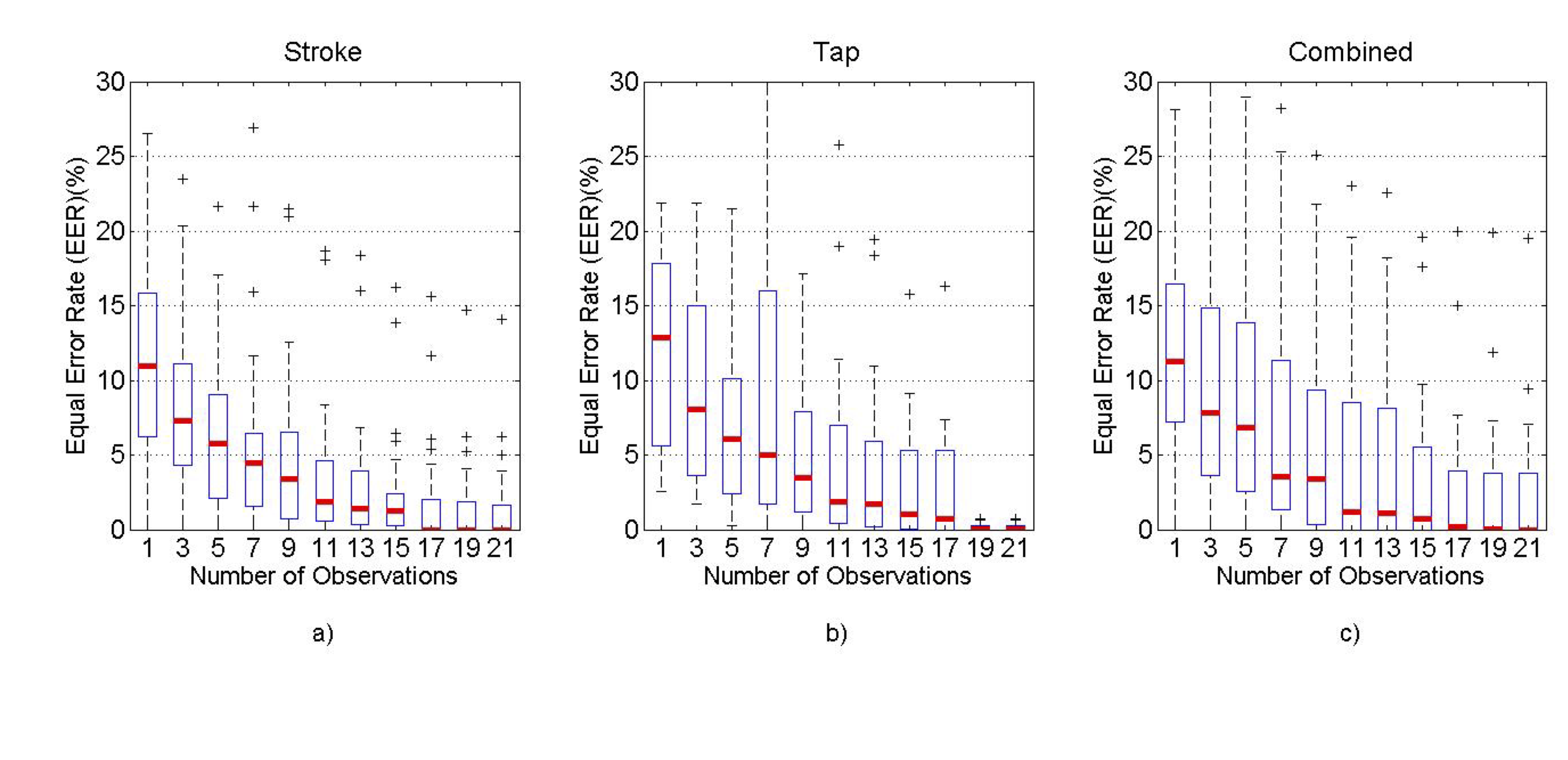}
\caption{Equal error rate variation as a function
of the number of gestures using multi-sensor information}
\label{w2eer}
\end{figure*}

The performance of the CHAS framework has been evaluated from the
two most frequently used gestures on mobile phone - tap and slide.
In this section, we describe the authentication results using only
touch information. The touch data obtained from the first two
tasks described in Section \ref{sec:Data_Set_Description} was used
to train the classifiers and the data from the following three
tasks was used for testing.

\textbf{EER Performance:} Figure \ref{wo2eer} presents the EER
performance of the CHAS framework. Since the number of consecutive
gestures required to get the final classification result reflects
the time requirement of the whole system, EER variation with the
number of observations has been plotted here. On each box, the
heavy red horizontal bar is the median, the edges of the box are
the first and third quartile, the whiskers extend to the most
extreme data points not considered outliers, and the outliers are
plotted individually.

Figure \ref{wo2eer}(a) shows that the median EER of authentication
using one observation of slide is 13.29\%. The EER decreases to
0\% after observing 19 consecutive slide gestures. Similarly, for
tap gesture, the median EER is found to be 16.55\% using one
observation. After 17 consecutive taps the EER saturates to near
1\% (see Figure \ref{wo2eer}(b)). Since slide gesture is
comparatively longer than tap, more data points are recorded for
slide.
The increased number of data points made the slide gesture more
distinguishable than tap. Thus, the authentication performance
based on slide model is better than the tap.

Next, we evaluate the performance of CHAS framework in a more
general scenario where the action sequences are random combination
of any types of gestures. Experimental results show that the
median EER is 15.10\% with one observation (see Figure
\ref{wo2eer}(c)). However, after 21 observations it is reduced to
0\%. Thus, it may be noted that when a combination of slide and
tap gestures is used, the EER becomes worse than using slide
gesture alone while better than only tap gesture.

\begin{figure}[!ht]
\centering
\includegraphics[height=4 cm,width=8 cm]{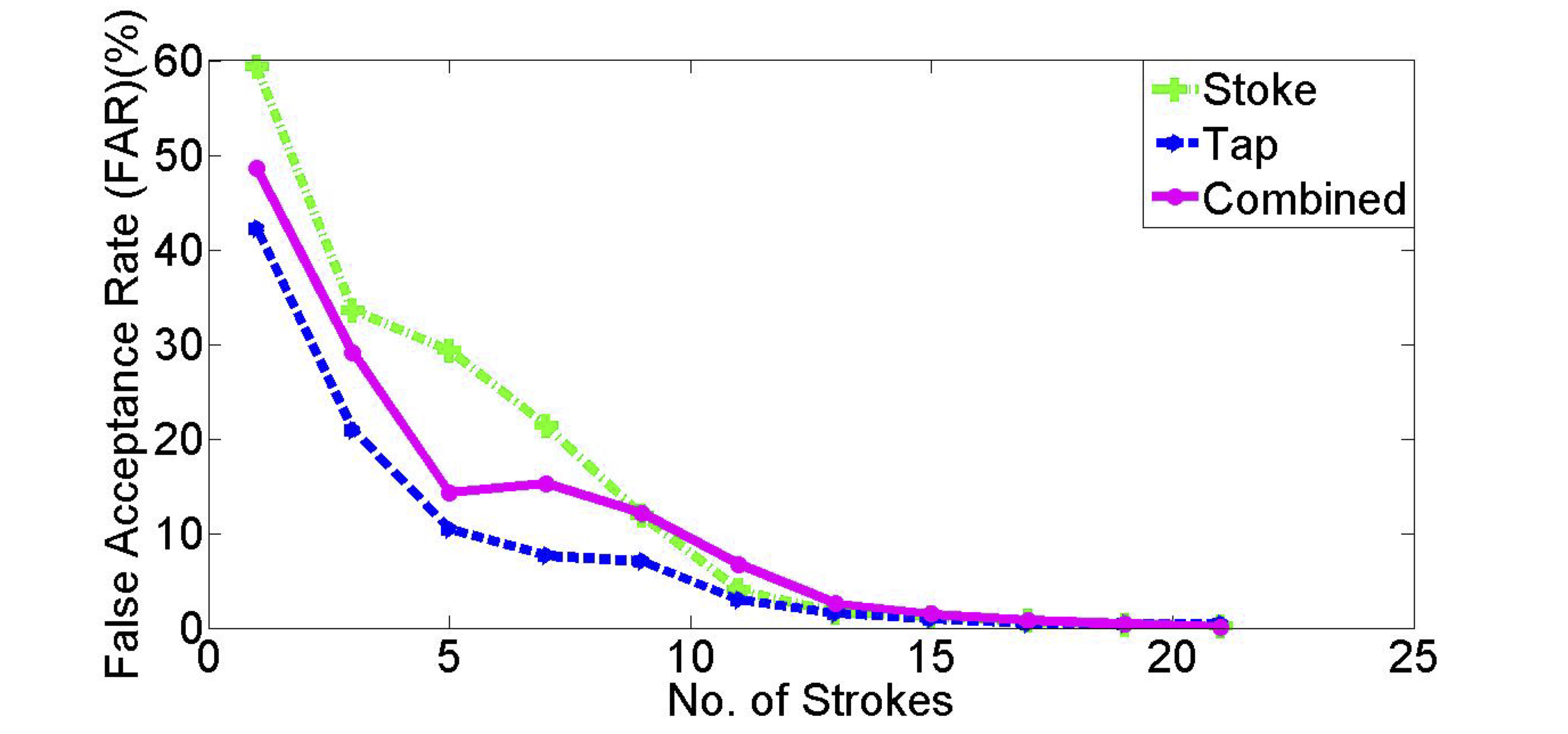}
\caption{False acceptance rate variation as a
function of the number of gestures using touch-pad input}
\label{wo2far}
\end{figure}

\textbf{FAR and FRR Performance:} We also look at the security and
usability of the proposed approach in terms of FAR and FRR. In
applications where security is not so much of importance (like
games), low FRR is desired. Figure \ref{wo2far} depicts the median
FAR performance of the system. When the FRR of the CHAS framework
is zero, the median FAR of the authentication is near 0\% after
observing 15 taps and 17 slide gestures.

\begin{figure}[!ht]
\centering
\includegraphics[height=4 cm,width=8 cm]{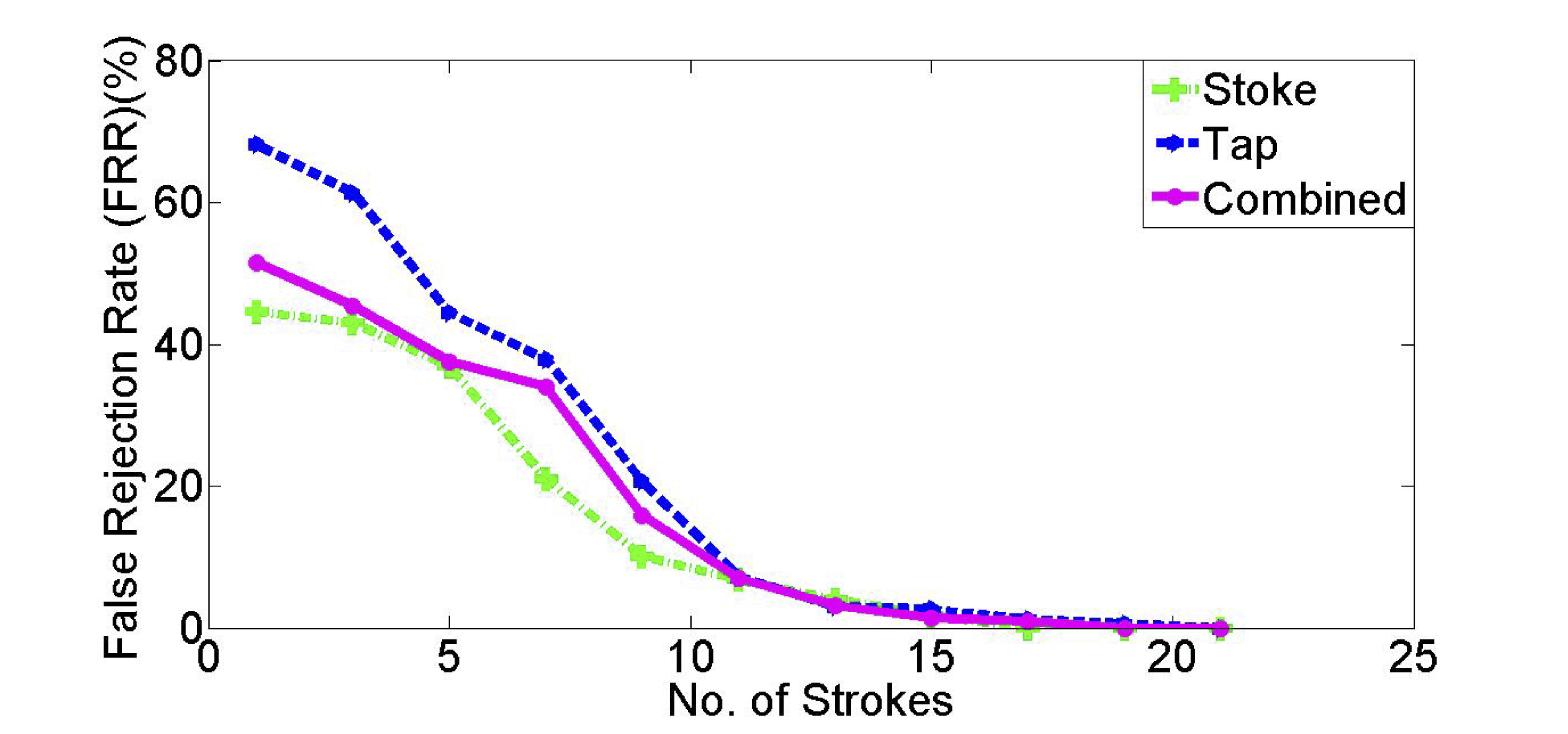}
\caption{False rejection rate variation as a
function of the number of gestures using touch-pad input}
 \label{wo2frr}
\end{figure}

For applications with high security requirement (like banking),
FRR performance of the CHAS framework is also evaluated while the
FAR is set to zero
(see Figure \ref{wo2frr}). It can be observed that the FRR reaches
0\% after 19 consecutive taps and 17 consecutive slide gestures.

Using combined gesture data, both the FAR and FRR come down to 0\%
after 17 gesture observations. These results indicate inherent
strength of the HMM based behavior model in varied usage
scenarios.

\subsection{Authentication with Multiple Sensor Data}

In this section, we evaluate the CHAS framework using touch as
well as two other sensors data. Similar to the former case, here
also training is done based on the touch, accelerometer and
gyroscope data obtained from the first two tasks and the rest of
the data is used for testing.

\textbf{EER Performance:} First, we show how the system works with
either tap or slide gesture (similarly to the previous
evaluation). Then, the system performance with general gesture
data set that features any combination of these two gestures is
reported.

Figure \ref{w2eer} presents the EER performance of the two
gestures with increasing number of observations. It can be seen
that the median EER
using one observation of tap is 12.85\% and one slide is 10.92\%.
After observing about 19 consecutive taps and 17 consecutive slide
gestures, the EER is reduced to 0\%. However, while using combined
gestures, the EER is found to be 11.25\%. It decreases to 0\%
after
17 consecutive gestures. Therefore, it may be noted that inclusion
of the multi-sensor data helps to improve the performance,
resulting in lower EER compared to using only the touch data for
the same number of consecutive gestures.

These results are comparable to the results achieved using only
vertical or horizontal slide data \cite{roy2014hmm, Frank13},
where 11 consecutive strokes were needed to achieve near 0\% EER.
Thus, with a small increase in the number of required gestures,
CHAS framework achieves similar accuracy while handling almost
every possible gesture performed on a mobile device in
unconstrained manner. Moreover, since this result is achieved with
a combination of gestures where tap generally requires less time
than slide gesture, overall time requirement to get the
verification result of the CHAS framework is close to
\cite{roy2014hmm}.

\begin{figure}[!ht]
 \centering
\includegraphics[height=4 cm,width=8 cm]{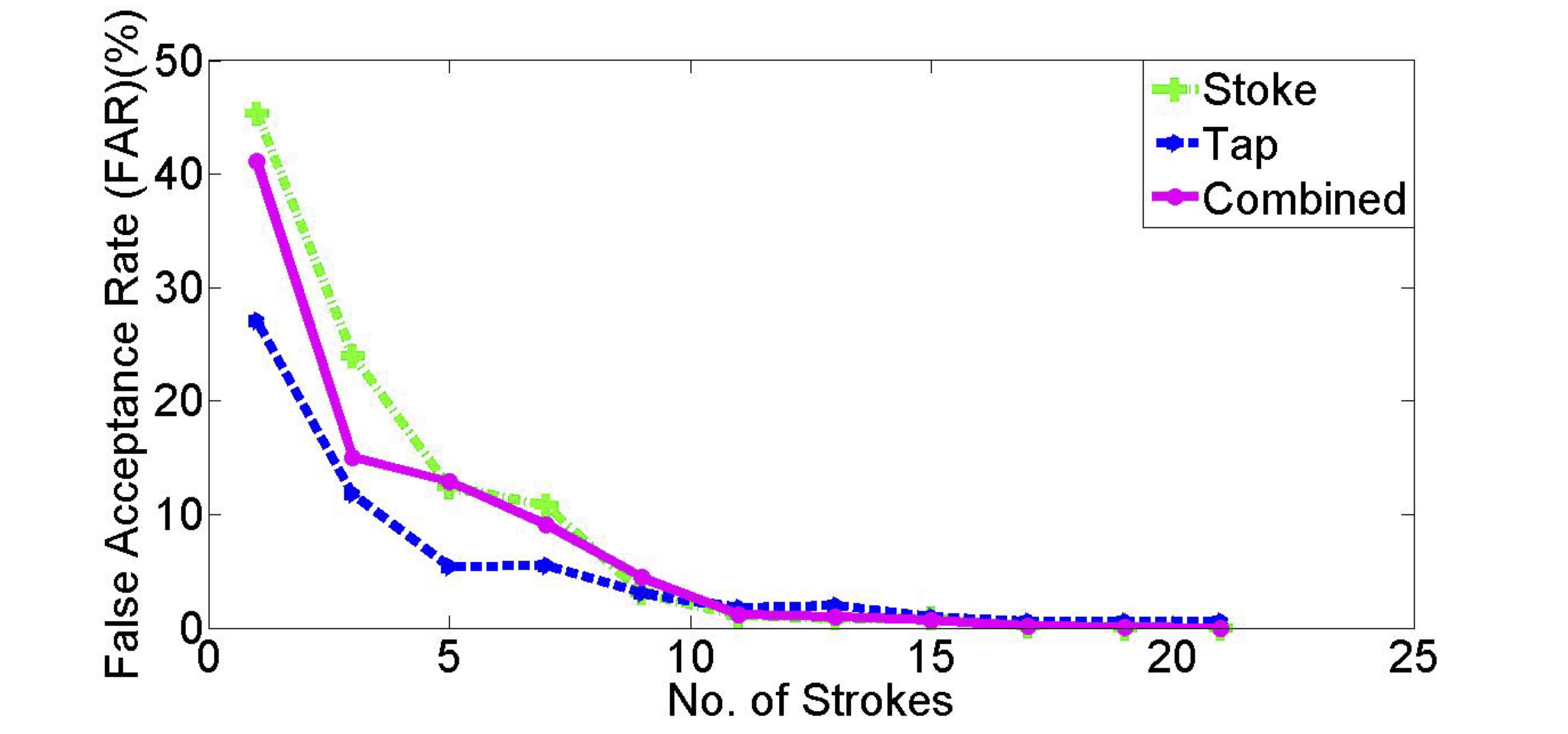}
\caption{False acceptance rate variation as a
function of the number of gestures using multi-sensor information}
\label{w2far}
\end{figure}

\begin{figure}[!ht]
 \centering
\includegraphics[height=4 cm,width=8 cm]{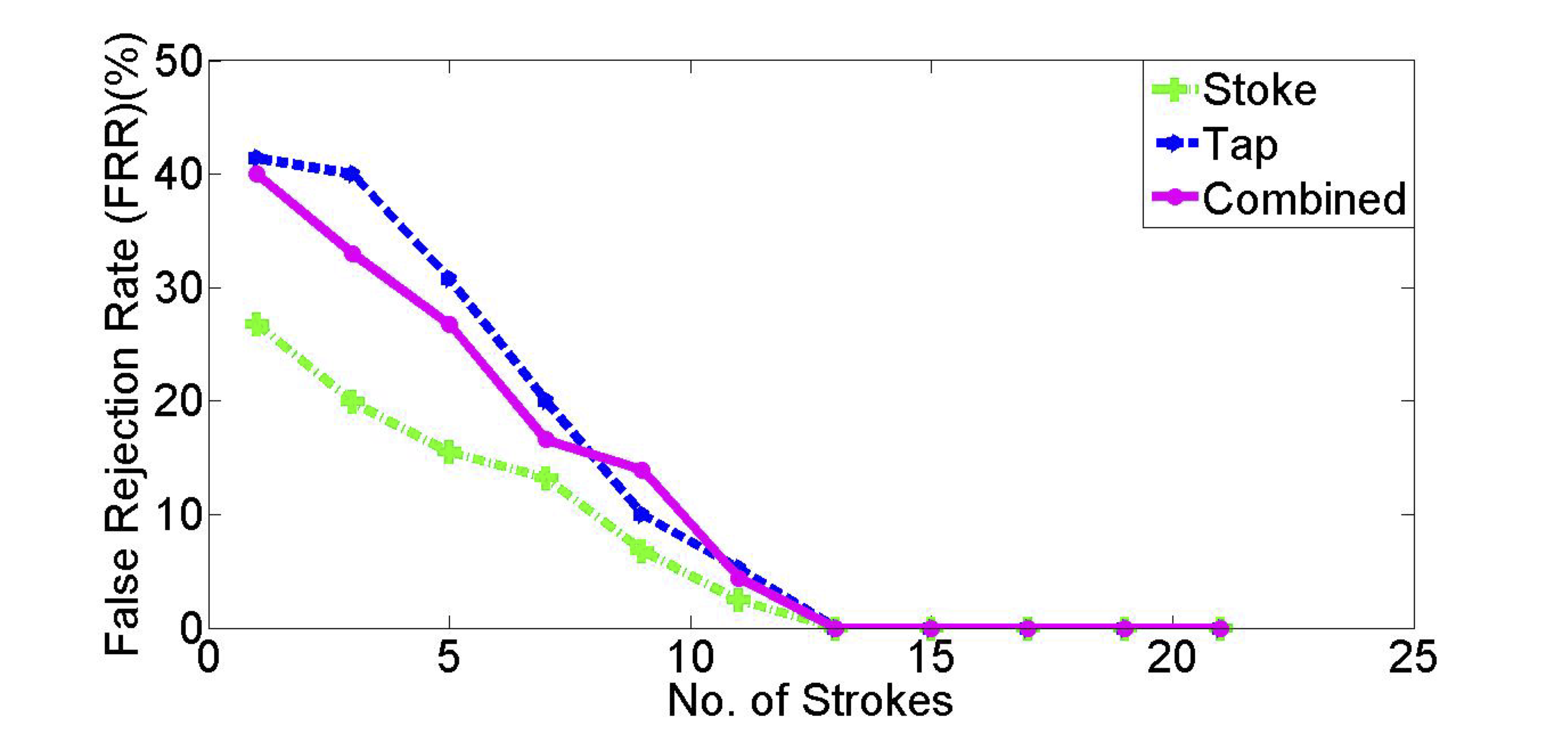}
\caption{False rejection rate variation as a
function of the number of gestures using multi-sensor information}
\label{w2frr}
\end{figure}

\textbf{FAR and FRR Performance:} Figure \ref{w2far} presents the
FAR performance of the system. The results show that the median
FAR reaches 0\% after observing about 15 consecutive taps and 13
consecutive slide gestures. Similarly, the median FRR decreases to
0\% with 13 consecutive taps or slide gestures (see Figure
\ref{w2frr}). Using combined gesture, the FAR decreases to 0\%
after 11 consecutive observations, while the FRR decreases to the
same level with 13 gestures. Thus, the FAR and FRR performance of
CHAS framework also improved after including multi-sensor data
with touch information.

\section{Conclusions}

This work presents a survey that points out the need for enhanced
security mechanism beyond the point of traditional login. This
observation motivated us to design CHAS framework that models user
touch behavior using HMM. The authentication method is based on
multi-sensor data recorded from the owner's touch interactions on
his mobile device. Extensive evaluation of the proposed approach
on a newly acquired database, which features real-life mobile
usage scenario, has been carried out.

The first part of the work demonstrates that it is possible to
authenticate users with reasonable accuracy using only the touch
data of the owner. The benefits of using only the device owner's
data are twofold. First, in case of personal devices, data from
other users may not be available. Thus, training the classifier
with other users' data is not possible. Second, authentication
results using only the owner's data for training reflect the real-life situation in
a better way.

Next, the work continues by integrating additional data from
accelerometer and gyroscope sensors. The users were not restricted to do
any specific gestures on the phone but were allowed to perform
different activities freely. Testing on the new acquired data set
shows that the proposed framework achieves robust authentication
result, with 0\% ERR using
17 consecutive observations.

The results suggest that the CHAS framework is applicable in a
variety of situations, both using only touch data as well as
multi-sensor data, to execute continuous authentication based on
natural touch interactions. 
Future work involves automatic update of the CHAS framework with
new data over time without the need of retraining.

\section*{Acknowledgment} This material is based in part upon work supported by the National
Science Foundation under Grant No. 1228842. We would like to thank
T.V. Nguyen for sharing some results of his mobile security survey
with us.

\bibliographystyle{abbrv}

%

%
%
%
%
%
%

\end{document}